\documentclass[journal]{IEEEtran}
\usepackage{amsmath}
\usepackage[pdftex]{graphicx}
\usepackage{algorithmic}
\usepackage{algorithm}
\usepackage{array}
\usepackage{url}
\usepackage{stfloats}
\usepackage{booktabs}
\usepackage{multirow}
\usepackage{bm}
\usepackage[super]{nth}

\if CLASSOPTIONcompsoc
  \usepackage[caption=false, font=normalsize, labelfont=sf, textfont=sf]{subfig}
\else
  \usepackage[caption=false, font=footnotesize]{subfig}
\fi

\if CLASSOPTIONcaptionsoff
  \usepackage[nomarkers]{endfloat}
 \let\MYoriglatexcaption\caption
 \renewcommand{\caption}[2][\relax]{\MYoriglatexcaption[#2]{#2}}
\fi

\begin{document}
\title{Adaptive Multi-scale Detection of Acoustic Events}

\author{Wenhao~Ding,~\IEEEmembership{Student Member,~IEEE,} and~Liang~He,~\IEEEmembership{Member,~IEEE}%

\thanks{W. Ding and L. He are with the Department of Electrical Engineering, Tsinghua University, Beijing, 100084 China. L. He is the corresponding author. E-mail: dingwenhao95@gmail.com, heliang@tsinghua.edu.cn}%

\thanks{Manuscript received April 24, 2019; revised August 26, 2019.}}

\markboth{IEEE/ACM TRANSACTIONS ON AUDIO, SPEECH, AND LANGUAGE PROCESSING, VOL. X, NO. X, MARCH 2019}%
{Wenhao \MakeLowercase{\textit{et al.}}: Adaptive Multi-scale Detection of Acoustic Events}

\IEEEpubid{0000--0000/00\$00.00~\copyright~2019 IEEE}

\maketitle

\begin{abstract}
The goal of acoustic (or sound) events detection (AED or SED) is to predict the temporal position of target events in given audio segments. This task plays a significant role in safety monitoring, acoustic early warning and other scenarios. However, the deficiency of data and diversity of acoustic event sources make the AED task a tough issue, especially for prevalent data-driven methods.
In this paper, we start from analyzing acoustic events according to their time-frequency domain properties, showing that different acoustic events have different time-frequency scale characteristics.
Inspired by the analysis, we propose an adaptive multi-scale detection (AdaMD) method.
By taking advantage of hourglass neural network and gated recurrent unit (GRU) module, our AdaMD produces multiple predictions at different temporal and frequency resolutions.
An adaptive training algorithm is subsequently adopted to combine multi-scale predictions to enhance the overall capability.
Experimental results on Detection and Classification of Acoustic Scenes and Events 2017 (DCASE 2017) Task 2, DCASE 2016 Task 3 and DCASE 2017 Task 3 demonstrate that the AdaMD outperforms published state-of-the-art competitors in terms of the metrics of event error rate (ER) and F1-score.
The verification experiment on our collected factory mechanical dataset also proves the noise-resistant capability of the AdaMD, providing the possibility for it to be deployed in the complex environment.

\end{abstract}


\IEEEpeerreviewmaketitle

\section{Introduction}

\IEEEPARstart{T}{he} task of detection and classification of acoustic events has attracted much attention in recent years \cite{70, 7, 61}. The classification task is to determine the category, while the detection task is to predict the temporal position of the target events. Obviously, the latter one provides us more information: accurate positions of target events not only enable us to find meaningful fragments in the tedious background audio, but can also be utilized in multi-modal analysis with other synchronous sensors (e.g. video, radar, lidar and \emph{etc}.) to obtain more comprehensive information \cite{71, 72}. In light of these advantages, audio event detection (AED) system has been applied in many practical fields, including detection of abnormal sounds in the transportation \cite{26, 27}, detection of violence in monitoring system \cite{31}, and detection of events in indoor environment such as meeting room or house \cite{30, 32}.

However, another reason why AED task deserves much attention is that it is still a problem far from being solved \cite{16}. We believe the difficulty mainly originates from three characteristics:

\begin{enumerate}[]
\item \textbf{Data is extremely unbalanced.} Compared with background segments, the length of one acoustic event is limited. Since positive frames are few, artificial synthesis method is usually used to generate training data by inserting acoustic event sources into a large number of backgrounds. However, repeated positive samples can easily lead to over-fitting for data-driven methods.

\item \textbf{Events have diverse characteristics.} In most situations, different acoustic events have different unique patterns. We have to take it into consideration when designing our models rather than using a general model for all kinds of acoustic events.

\item \textbf{Time-frequency scale is not consistent.} The variation in duration for acoustic events is relatively large, Therefore, analyzing and extracting features under a fixed scale by splitting them into segments could be ineffective.
\end{enumerate}

\IEEEpubidadjcol

\begin{figure*}[!t]
\centering
\includegraphics[width=18cm]{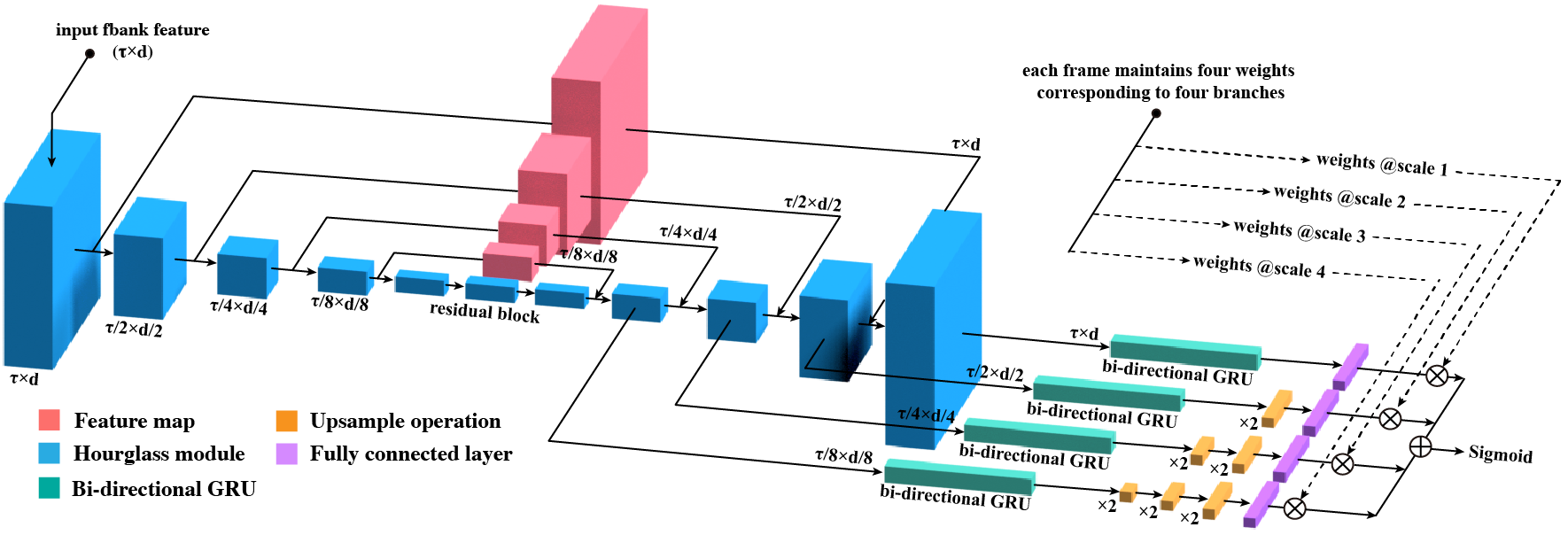}
\caption{Scheme of AdaMD, consisting of three parts: hourglass module (blue), bi-directional GRU modules (orange) and upsampling modules (green). $\tau$ is the dimension of time axis, and $d$ is the dimension of feature. The hourglass module contains one residual block in each layer. The details of the adaptive training process is summarized in Algorithm~\ref{alg1}.}
\label{structure}
\end{figure*}

To deal with the aforementioned difficulties, we propose an adaptive multi-scale detection method (AdaMD), whose structure is shown in Fig.~\ref{structure}. In view of the excellent performance of convolutional recurrent neural network (CRNN) \cite{75} in processing time sequences \cite{1, 2}, our high-level structure consists of a CNN and a RNN network. In the CNN part, we use a network structure called \textit{hourglass} \cite{9}, which has been widely used in keypoint detection tasks of computer vision field. The advantage of this structure comes from the capability of extracting features with multiple time-frequency resolutions. In the RNN part, we adopt the gated recurrent unit (GRU) module \cite{74} in each branch to process temporal information. Then, connected with the upsampling layer, each branch outputs a prediction that has the same length as the input sequence. During the training, we borrow the idea from AdaBoost \cite{55} that regards multi-scale branches as multiple weak classifiers.
Since the time-frequency scales of audio events are inconsistent, each branch may only be suitable for dealing with samples of one certain scale. Therefore, in order to maximize the performance of each classifier, we will weaken the influence of the branch that outputs the worst result.

In order to compare our system with related works, we firstly conduct experiments on DCASE 2017 Task 2 Dataset \cite{2}, the most commonly used for rare acoustic event detection. Our results on both development and evaluation dataset outperform currently available methods on this dataset, including the first place \cite{1} in the competition.
To extend this method to the non-rare acoustic event detection, we also test AdaMD on DCASE 2016 Task 3 \cite{70} and DCASE 2017 Task 3 \cite{7} and achieve better results than the first place in both datasets \cite{81}\cite{82}. In addition, we carry out a verification experiment in real-world environment, which requires detecting mechanical failure events of factory machine. The promising results demonstrate that our method is able to deal with AED tasks even in much complex environments.

In summary, our contributions are three-fold:
\begin{enumerate}[]
\item We provide an analysis for acoustic events about characteristics in time-frequency domain.
\item We develop an adaptive multi-scale detecting method that processes acoustic events with different resolutions.
\item We achieve better performances than other state-of-the-art methods on DCASE 2016 and DCASE 2017 datasets.
\end{enumerate}

The rest of this paper is organized as follows. Three related tasks are briefly discussed in Section \uppercase\expandafter{\romannumeral2} and potential solutions for AED tasks are introduced in \uppercase\expandafter{\romannumeral3}. Then, we describe our analysis of acoustic events in Section \uppercase\expandafter{\romannumeral4} and our adaptive multi-scale detector in Section \uppercase\expandafter{\romannumeral5}. The experimental results and discussions are presented in Section \uppercase\expandafter{\romannumeral6} followed by the conclusions in Section \uppercase\expandafter{\romannumeral7}.

\section{Related Tasks}

In this section, we discuss three kinds of related tasks, i.e. polyphonic event detection, weakly supervised event detection and anomaly event detection. All of them attract much attention in acoustic event detection field as well.

\subsection{Polyphonic Event Detection}

In general, there are two situations in AED tasks. One is that there is only one or zero event in a segment, which is called monophonic detection \cite{21, 22, 23}. The other one is that multiple events appear at the same time in one segment with overlaps. The model is required to predict not only the onset and offset, but also the type of the event. This is called polyphonic detection \cite{18, 19, 20}. Intuitively, training with multiple labels can leverage more information, and it has been proved that polyphonic method can achieve better results on polyphonic detection tasks \cite{76}. Our method is not only designed for monophonic detection, since it can be extended by using multiple fully-connected layers to predict the probability for multiple events.

\subsection{Weakly Supervised Event Detection}

Weakly supervised method is also a common kind of algorithm for AED tasks \cite{34, 35, 36}. Usually, it is time-consuming and laborious to accurately annotate the onset and offset for one acoustic event. In contrast, it is much easier to only annotate the category. When category annotations are the only thing available, while the temporal position are required from the model, we are dealing with the weakly label event detection. This task is included in both DCASE 2017 and DCASE 2018 competitions. To tackle with this problem, \cite{34} utilized the gated linear unit (GLU) \cite{62} to replace the ReLU \cite{58}, and won the championship of DCASE 2017 Task 4, while the champion of DCASE 2018 Task 4 was achieved by \cite{36} with a teacher-student network structure. Both of them aim at extracting position information from the bottleneck layer. Although training resources under weak supervision are relatively easy to obtain, this direction is still in its infancy. Even the best systems can't achieved satisfactory results, compared with supervised counterparts.

\subsection{Anomaly Event Detection}

One characteristic of AED task is the ratio of concerned acoustic event to the background is quite small, which is similar to the anomaly event detection task \cite{67, 68}. The difference is that we already know the types of audio events in AED tasks \cite{64}, so we need to look for events with known features, but in anomaly event detection we have no information about the abnormal samples to be discriminated.

For anomaly event detection task, \cite{64} proposes a framework: firstly, they construct a complete set in a manifold space, then subtract the normal part from the complete set to obtain information about anomalous audio and recognize it. Furthermore, \cite{69} uses the Kullback-Leibler (KL) divergence to measure the similarity between normal and abnormal samples according to the short-time Fourier transform (STFT) feature. In addition, \cite{66} resorts to a variety of front-end feature fusion (FFT, DCT and MFCC). As mentioned earlier, the anomaly detection method is generally target-less, which means it is only applicable to the case of simple background. When the environment gets complex, false alarm is hard to avoid. On the other hand, for AED tasks, we do have the information of the target events, but the anomaly events detection does not utilize this information.

\section{Potential Solutions}

In this section, we present a description for four kinds of potential solutions for AED task: traditional signal processing methods, CRNN method, region proposal method and multi-scale method.

\subsection{Traditional Signal Processing Method}

In the early stage, traditional signal processing methods with shallow model backend have been tried to solve the AED tasks. The representative works are listed here: \cite{23} proposes a random forest approach to predict the distance between frames in acoustic event segments to determine the position of onset and ending; \cite{37, 38} use non-negative matrix factorization (NMF) to detect target acoustic event; \cite{39, 45} use support vector machine (SVM) to classify audio events; \cite{47} constructs a gaussian mixture model (GMM) for source events and background respectively, and then merges them into a joint GMM for event position detection; \cite{46, 21} construct a hidden Markov model (HMM) for separation.

Some of these methods attempt to directly predict onset and offset of audio events, while others try to separate acoustic events from background sounds. However, the common problem of traditional methods is the deficiency of data utilization, leading to mismatch model that can not extract highly relevant features for target events.

\subsection{CRNN Method}

In order to make better use of the limited data, current mainstream researches prefer data-driven methods with deep neural networks. Some use CNN \cite{3, 40} or RNN \cite{19, 41} separately, but only achieve few promising results. CNN has a strong ability to extract local features, while RNN structure is capable to obtain more information by processing time sequences. Therefore, combining these two structures is expected to achieve better results. At present, CRNN \cite{75} has gradually become the state-of-the-art structure to deal with AED tasks \cite{1, 2, 20, 42, 43, 44}. For example, \cite{1} got the first place in DCASE 2017 Task 2 with the combination of a 1D-convolution structure and a bi-directional GRU structure. \cite{43, 44} combined attention mechanism with RNN, making the region of target events better concerned during feature extraction.

\subsection{Region Proposal Method}

Inspired by state-of-the-art detection methods in computer vision field \cite{63}, region-based method has been widely applied to AED tasks. This method is different from previous ones, since it directly provides the onset and offset with an end-to-end pipeline: firstly, it gives a large number of possible candidate positions with different scales, and then selects the final prediction with the highest probability. For example, the structure of R-FCN \cite{63} was used in \cite{4}, and a better framework called faster-RCNN \cite{6} was utilized in \cite{5}. However, the problem is similar to that of computer vision field: the parameters of neural network need to be pre-trained in advance, otherwise it will be difficult to converge. Although Mask-RCNN \cite{80} might be a solution to avoid the pre-train operation, this method has not been used in acoustic event detection task yet.

\subsection{Multi-scale Method}

\cite{54, 8} are typical works that apply the concept of multi-scale feature extraction in AED tasks. However, the multi-scale in \cite{8} is only reflected from the different size of convolution kernels, which is a relatively simple structure, without considering the fusion of multiple scales. Although \cite{54} extracts features from different temporal scales and combines them, the hierarchic structure makes the training process quite slow.

Besides, multi-scale feature extraction in frequency level is also important \cite{77}. Intuitively, different frequency ranges dominate different acoustic events, which makes it hard to predict accurate results only with certain frequency range.

In fact, the application of multi-scale methods are widely concentrated in computer vision, since images usually possess multi-scale information. In the keypoint detection task of human body, \cite{9} proposes a structure named \textit{hourglass} that reduces the dimension and then increases the dimension of an image, and retains the previous information during the dimension increasing process. As this structure has achieved good results, other tasks begin to benefit from it \cite{10, 11, 12}. Recently, a lot of new methods has been proposed to improve the basic structure of \textit{hourglass} \cite{13, 14, 15}. In audio related filed, there are also works trying to introduce multi-scale structure \cite{78}\cite{79} and achieve promising results.
Inspired by \cite{9}, our AdaMD takes \textit{hourglass} as the feature extractor, which performs better than \cite{8} and \cite{54}.

\begin{figure*}[!t]
\centering
\includegraphics[width=18cm]{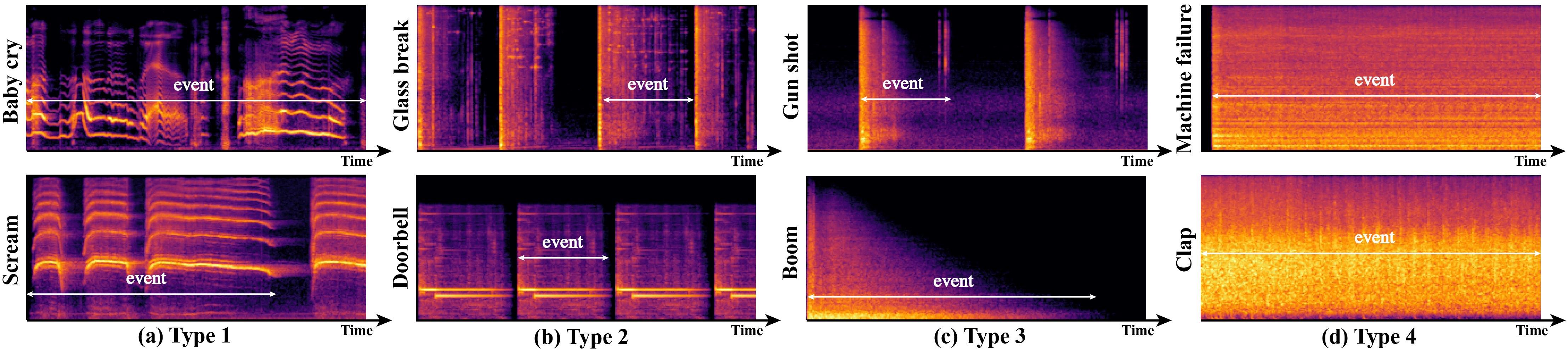}
\caption{Real-life acoustic event examples. We roughly divide them into four types and each column represents one category. This may not be a precise classification since many events can not be classified in one of these categories. The white label represents the duration of the event.}
\label{taxonomy}
\end{figure*}

\section{Analysis for Acoustic Events}

In this section, we analyze the characteristics of acoustic events in DCASE 2017 Task 2 from two perspectives, and then propose to roughly divide them into four categories. We will also discuss how this analysis inspire our multi-scale detection algorithm.

\subsection{Characteristic Analysis}

Through our research experiences, we find that the onset of one acoustic event is usually apparent, while the ending position is relatively vague. One possible explanation is that acoustic events are usually instantaneous, some of which have sudden terminations (e.g. \textit{baby cry}), while others have vanishing terminations (e.g. \textit{gun shot}). In view of this feature, we can firstly classify events based on whether there is a vanishing tail at the ending position.

The existence of vanishing tail makes annotation quite difficult. If the entire event segment is annotated with the same label (e.g. 0/1 label), the model will be force to learn features of both the onset and the ending position with the same label. If soft label (e.g. a value from 0 to 1) is used, all frames need to be annotated according to the different degree of tailing of the event, which requires much labor. On the other hand, there is no annotation problem if the event ends abruptly.

The second classification is based on the consistency of features. During the occurrence of an event, the feature may be inconsistent (e.g. \textit{baby cry}) or consistent (e.g. \textit{glass break}). The consistency of features here is different from the vanishing tail mentioned above. Take \textit{glass break} for example, this event has a vanishing tail since it has a decrease in intensity, but the features inside the class of \textit{glass break} are quite similar. In contrast, \textit{baby cry} and other events related to human vocal system have high intra-class variations, which is concluded by inconsistent features.

The lack of consistency makes feature extraction difficult. For inconsistent features, we need to analyze them from different levels, otherwise, it will be difficult for the subsequent recognition process. We provide the f-bank features of two real-life examples for each category in Fig.~\ref{taxonomy}.

\begin{table}[th]
\begin{center}
\caption{Description of Four-category Acoustic Events}
\label{taxonomy_table}
		\begin{tabular}{c||c|c|c}
		\hline
		Type  & Tail & Feature & Examples \\
		\hline
		1 & non-vanishing & inconsistent & baby cry, scream \\
		\hline
		2 & vanishing & consistent & glass break, doorbell \\
		\hline
		3 & vanishing & inconsistent & gun shot, boom \\
		\hline
		4 & non-vanishing & consistent & machine failure, clap \\
		\hline
		\end{tabular}
\end{center}
\end{table}

\subsection{Insights to Model Design}

It is obvious that the acoustic events are quite different in both time and frequency domains. These characteristics make the events hard to be handled within a single scale. Even if the metric in DCASE 2017 Task 2 only considers the onset accuracy, most of the deep learning methods need to use the frame-wise label for learning the active probability of every frame. That means, the characteristics we summarized will also influence the result in DCASE 2017 Task 3. Therefore, above analysis given us an inspiration to develop a multi-scale algorithm for AED tasks.

\section{Adaptive Multi-scale Detector}

In this section, we first introduce the general structure of our AdaMD method, then describe the details of the \textit{hourglass} structure, and finally explain the adaptive multi-scale fusion method inspired by AdaBoost algorithm \cite{55}.

\subsection{General Structure}

\begin{figure}[!t]
\centering
\includegraphics[width=8cm]{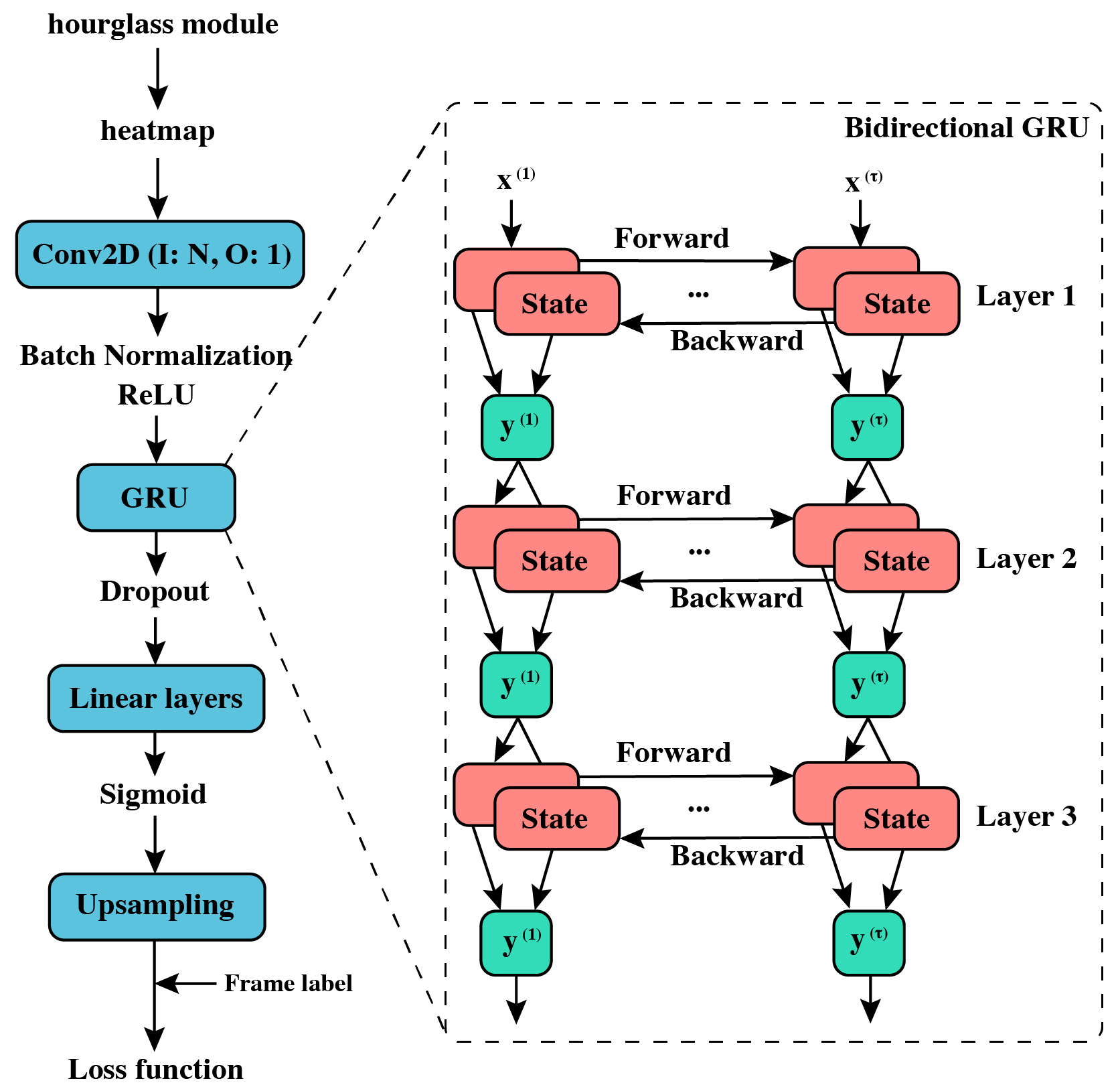}
\caption{Structure of one branch (weak classifier). The input of this part is the output of hourglass, whose dimension is $(\tau, d, N)$. The first 2d-convolution layer is used for reducing the channel from N to 1. The GRU block contains a bidirectional GRU module, whose details are shown in the right of the figure.}
\label{crnn}
\end{figure}

Our method adopts the prevalent CRNN \cite{75} as the basic structure. We utilize \textit{hourglass} instead of common CNN layers based on our above analysis. Then, the bi-directional GRU module is cascaded to the outputs in each scale of the hourglass. The advantage of bi-directional GRU module \cite{74} is the capability to utilize forward and backward information of time sequences to obtain features that the uni-directional counterpart does not have. Since the output's dimensions of all scales are different, the dimension of the input layer and the hidden layer of these GRU modules are also different. The dimension of input layers and hidden layers are [16, 32, 64, 128] and [32, 32, 64, 64] respectively, and the number of hidden layer is 3. Some Details are shown in Fig.~\ref{crnn}.

As for the data pipeline, we regard the data we input into the neural network as samples, whose dimension is $\tau\times d$, where $d$ is the dimension of Mel feature and $\tau$ is the length in time axis. We clip one audio file into multiple samples during training, while in validation and testing stage, we regard one audio file as one sample and input the whole sequence into the model. That's because the initial hidden state of GRU is empty, which will make the prediction for the first few frames unpredictable.

In the loss function part, we use frame-wise label that has the same length as the input sequence, thus for each branch we conduct upsampling operation to make the final output length equal to the length of label. After that we can directly calculate the loss for all branches.

In our experiments, we choose a four-layer hourglass model (the same as \cite{9}), so there will be four terms in the loss function. A simple idea is to directly use the average of these terms as the final output probability. However, this will corrupt the correct results in one certain scale if other branches are wrong. As a remedy, we propose an adaptive fusion method in next subsection to maximize the overall accuracy.

\begin{figure}[!t]
\centering
\includegraphics[width=8cm]{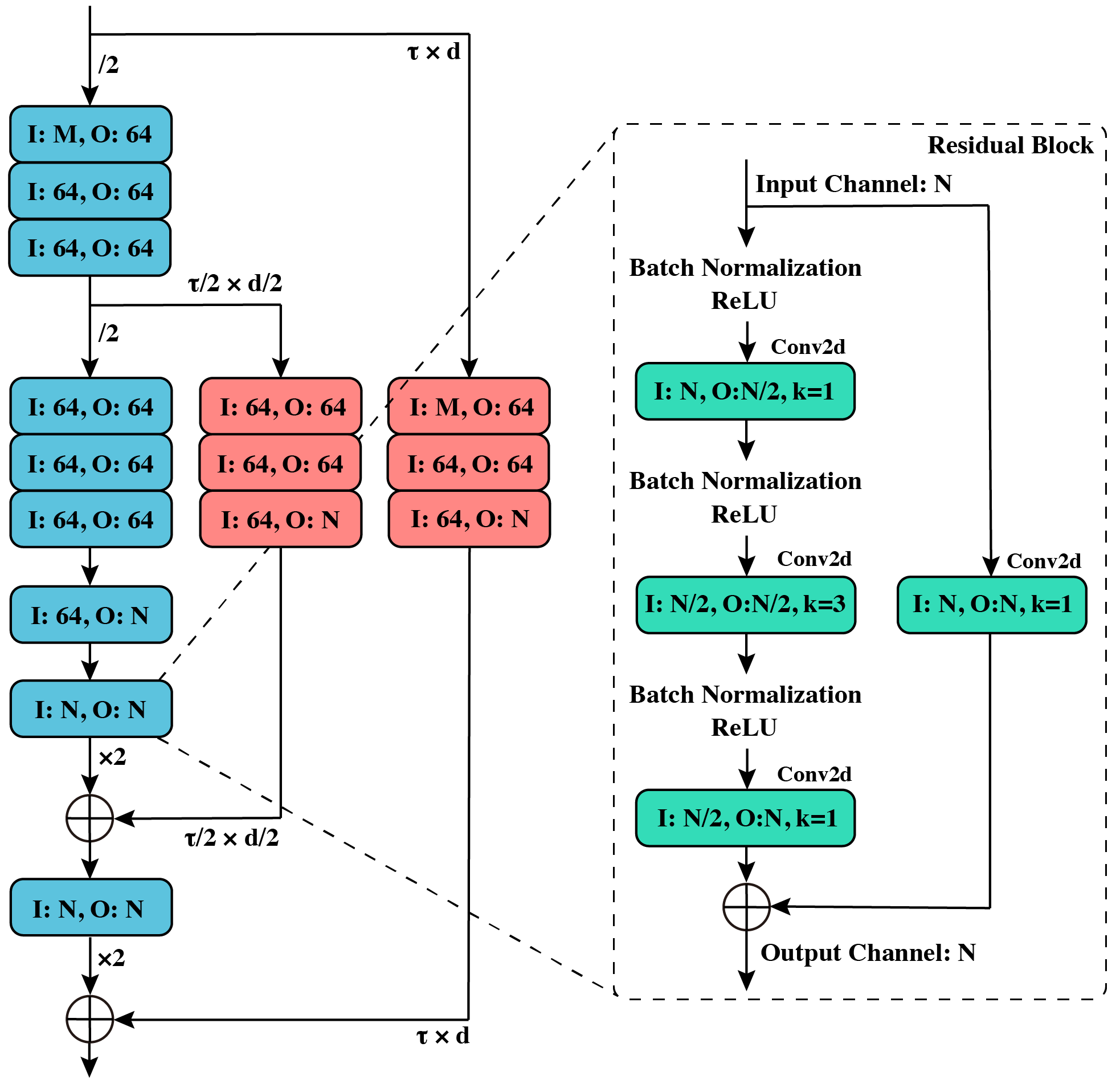}
\caption{Structure of a two-layer hourglass module. \textbf{I} means the number of input channel, \textbf{O} means the number of output channel, and \textbf{k} means the kernel size of convolution layer. The blue block represents a residual block, whose details are shown in the right of the figure.}
\label{hourglass}
\end{figure}

\subsection{Hourglass Network}

Hourglass network \cite{9}, whose name comes from the similar shape to an hourglass, is a novel structure used in keypoint detection tasks. It firstly reduces the original image to a low-dimensional heatmap through downsampling, and then restores the heatmap back to the original dimension through upsampling. In the recovery process, the information of the corresponding dimension in the downsampling process will be used. The final output is a set of multi-channel heatmaps with the same size as the original input, each of which contains the distribution of the keypoint. Inside of each scale, one residual module \cite{73} is used to avoid the gradient disappearance caused by deep network structure. The details of this pipeline is shown in Fig.~\ref{hourglass}.

There are two reasons for using this structure in AED tasks. First of all, AED tasks can be regarded as a 1-dimensional keypoint detection task, in which the target event can be seen as the keypoint in time domain. Thus, predicting the active frames (not only the onset and ending position) of event is similar to predicting the keypoints. Secondly, this multi-scale structure is used to solve the problem of inconsistent time-frequency scales mentioned in Section \uppercase\expandafter{\romannumeral1}. With hourglass network, we are able to obtain information from different resolutions (see Fig.~\ref{multiscale_missing}).

\subsection{Adaptive Fusion}

Inspired by the AdaBoost algorithm \cite{55}, we regard all GRU branches as weak classifiers. The objective is to force each classifier to deal with samples as well as possible. At the same time, there will be a weight to balance the contribution of all branches. In other words, if one branch learns some features that perfectly recognize the event, this branch should obtain more attention than others in the final output. In Adaboost algorithm \cite{55}, training samples are given weights according to the difficulty, while in our adaptive algorithm, we assign different weights according to the performance of samples in different scales. The weight of one sample for scale $k$ is:
\begin{equation}
s_k=
\begin{cases}
1, & if\ k = \mathop{\arg\min}_{k}\{l_k, i=1...K\}\\
\alpha, & else
\end{cases}
\end{equation}
where $l_k$ is the loss of $k^{th}$ branch of one sample, K is the number of branches. $\alpha$ is a hyper-parameter that control the contribution of worst branch. In order to determine the value of $alpha$, we search from 0.01 to 0.5 and set it to 0.1 for all datasets. Intuitively, this process will force one weak classifier give up samples that are too difficult for it, so that these samples will not interfere with the feature extraction process in this classifier. For the final output, we calculate four weights based on the accuracy of each weak classifier on the validation dataset, these weights are used to merge predictions of all branches:
\begin{equation}
w_k = \frac{v_k}{\sum_{k=1}^K v_k},\ \ \bm{p} = \sum\limits_{k=1}^K w_k\times \bm{{p}_k}
\end{equation}
where $v_k$ is the accuracy for $k$ scale on validation dataset, it is calculated by Binary cross entropy (BCE) of every frame in this sample. $\bm{p_k}$ is the prediction of one sample (a vector) for $k$ scale and $\bm{p}$ (a vector) is the final prediction.

\subsection{Training and Testing Process}

During the training stage, we optimize the entire network simultaneously without any fine-tune technology. The optimizer we use is Adam, the initialization method we use is normal distribution, and some important hyper-parameter of training is shown in Tab.~\ref{parameter_table}.
We choose \textit{Sigmoid} as the activation function of the final layer, which assures that the frame-wise probabilities are in the range of $[0, 1]$. Then we use binary cross entropy (BCE) as the loss function for each branch, and the entire loss is calculated by weighted average method:
\begin{equation}
L = \sum\limits_{k=1}^K \sum\limits_{n=1}^N -s^k_n[y_n\cdot log p_n +(1-y_n)\cdot log(1-p_n)]
\end{equation}
where $y_n$ is the frame-wise binary label for sample $n$ and $p_n$ is the final prediction for sample $n$.

During the test stage, the same process will be repeated and the only difference is that we input the whole audio file into the neural network. After obtaining the active probability of each frame, we will conduct a post-processing to find the first active frame according to a threshold. We first use a mean filter to the final prediction $\bm{p}$ to output a smoother prediction. The mean filter will use the mean value of three adjacent frames ($p^{i-1}, p^{i}, p^{i+1}$) to replace $p^i$. Then we compare each frame with the pre-defined threshold $\lambda$ to get the sample-wise prediction output. This fusion algorithm is detailedly summarized in \ref{alg1} and some important hyper-parameters about feature extraction and post-processing are shown in Tab.~\ref{parameter_table}.

\begin{algorithm}[t]
\caption{\textbf{Adaptive Fusion Algorithm}}
\begin{algorithmic}[1]
\label{alg1}
\STATE \textbf{Train:}
\STATE N is the number of all samples
\FOR{$n$ in (1, $N$)}
\STATE ${o}_{n,k} \Leftarrow$ Hourglass Network\ (${i}_n$), $k\in 1,..,K$
\STATE ${loss}_n = \sum_{k=1}^K s_{k,n} \times Loss({o}_{n,k}, {label}_n)$
\ENDFOR

\STATE \textbf{Validation:}
\STATE $v_k \Leftarrow$ Accuracy of scale $k$
\STATE $w_k = {v_k}/{\sum_{k=1}^K v_k}$
\STATE \textbf{Test:}
\STATE ${p}_k \Leftarrow$ Prediction of scale $k$
\STATE $p_{AdaMD} = \sum_{k=1}^K w_k\times{p}_k$
\end{algorithmic}
\end{algorithm}

\renewcommand\arraystretch{1.2}
\begin{table}[th]
\begin{center}
\caption{Important Hyper-parameters}
\label{parameter_table}
		\begin{tabular}{c||c||c}
		\hline
		\centering
		Notation & Value & Description\\
		\hline
		\begin{minipage}{0.01\textwidth}\centering $lr$ \end{minipage} &
   	    \begin{minipage}{0.05\textwidth}\centering 0.001 \end{minipage} &
		\begin{minipage}{0.25\textwidth}\centering learning rate \end{minipage} \\
		\hline
		\begin{minipage}{0.01\textwidth}\centering $E$ \end{minipage} &
   	    \begin{minipage}{0.05\textwidth}\centering 100 \end{minipage} &
		\begin{minipage}{0.25\textwidth}\centering number of epoch \end{minipage} \\
		\hline
		\begin{minipage}{0.01\textwidth}\centering $B$ \end{minipage} &
   	    \begin{minipage}{0.05\textwidth}\centering 45 \end{minipage} &
		\begin{minipage}{0.25\textwidth}\centering batch size \end{minipage} \\
		\hline
		\begin{minipage}{0.01\textwidth}\centering $d$ \end{minipage} &
   	    \begin{minipage}{0.05\textwidth}\centering 128 \end{minipage} &
		\begin{minipage}{0.25\textwidth}\centering number of the Mel filter \end{minipage} \\
		\hline
		\begin{minipage}{0.01\textwidth}\centering $\tau$ \end{minipage} &
   	    \begin{minipage}{0.05\textwidth}\centering 512 \end{minipage} &
		\begin{minipage}{0.25\textwidth}\centering length of one segment \end{minipage} \\
		\hline
		\begin{minipage}{0.01\textwidth}\centering $\tau_{step}$ \end{minipage} &
   	    \begin{minipage}{0.05\textwidth}\centering 0.8 \end{minipage} &
		\begin{minipage}{0.25\textwidth}\centering step size of $\tau$ \end{minipage} \\
		\hline
		\begin{minipage}{0.01\textwidth}\centering $\alpha$ \end{minipage} &
   	    \begin{minipage}{0.05\textwidth}\centering 0.1 \end{minipage} &
		\begin{minipage}{0.25\textwidth}\centering weight for the worst branch \end{minipage} \\
		\hline
		\begin{minipage}{0.01\textwidth}\centering $K$ \end{minipage} &
   	    \begin{minipage}{0.05\textwidth}\centering 4 \end{minipage} &
		\begin{minipage}{0.25\textwidth}\centering number of branches \end{minipage} \\
		\hline
		\begin{minipage}{0.01\textwidth}\centering $n_{fft}$ \end{minipage} &
   	    \begin{minipage}{0.05\textwidth}\centering 2048 \end{minipage} &
		\begin{minipage}{0.25\textwidth}\centering number of the FFT in F-bank \end{minipage} \\
		\hline
		\begin{minipage}{0.01\textwidth}\centering $\lambda$ \end{minipage} &
   	    \begin{minipage}{0.05\textwidth}\centering 0.5 \end{minipage} &
		\begin{minipage}{0.25\textwidth}\centering default binary threshold \end{minipage} \\
		\hline
		\begin{minipage}{0.01\textwidth}\centering ${t}_{win}$ \end{minipage} &
   	    \begin{minipage}{0.05\textwidth}\centering 0.04s \end{minipage} &
		\begin{minipage}{0.25\textwidth}\centering window size of F-bank extraction \end{minipage} \\
		\hline
		\begin{minipage}{0.01\textwidth}\centering ${t}_{hop}$ \end{minipage} &
   	    \begin{minipage}{0.05\textwidth}\centering 0.02s \end{minipage} &
		\begin{minipage}{0.25\textwidth}\centering step size of F-bank \end{minipage} \\
		\hline
		\end{tabular}
\end{center}
\end{table}

\begin{table*}[th]
\begin{center}
\caption{Comparison with other published methods on DCASE 2017 Task 2 (Error Rate $/$ F-score)}
\label{compare_table}
		\begin{tabular}{c|c|c|c|c|c|c|c|c}
		\hline
		\multirow{2}*{Method}
			~ & \multicolumn{4}{c|}{\textbf{Development Dataset}} & \multicolumn{4}{c}{\textbf{Evaluation Dataset}}\\
			\cline{2-9}
			~ & \begin{minipage}{0.07\textwidth}\centering baby cry\end{minipage} &
				\begin{minipage}{0.075\textwidth}\centering glass break\end{minipage} &
				\begin{minipage}{0.07\textwidth}\centering guns hot\end{minipage} &
				\begin{minipage}{0.07\textwidth}\centering average\end{minipage} &
				\begin{minipage}{0.07\textwidth}\centering baby cry\end{minipage} &
				\begin{minipage}{0.075\textwidth}\centering glass break\end{minipage} &
				\begin{minipage}{0.07\textwidth}\centering gun shot\end{minipage} &
				\begin{minipage}{0.07\textwidth}\centering average\end{minipage}\\
		\hline
		
		DCASE2017 Baseline  \cite{7}      & 0.67$/$72.0 & 0.22$/$88.5 & 0.69$/$57.4 & 0.53$/$72.7 & 0.80$/$66.8 & 0.38$/$79.1 & 0.73$/$46.5 & 0.64$/$64.1\\
		DCASE2017 \nth{1} place \cite{1}  & 0.05$/$97.6 & \textbf{0.01}$/$\textbf{99.6} & 0.16$/$91.6 & 0.07$/$96.3 & 0.15$/$92.2 & 0.05$/$97.6 & 0.19$/$89.6 & 0.13$/$93.1\\
		DCASE2017 \nth{2} place \cite{2}  & $-\ /\ -$   & $-\ /\ -$   & $-/-$       & 0.14$/$92.9 & 0.18$/$90.8 & 0.10$/$94.7 & 0.23$/$87.4 & 0.17$/$91.0\\
		DCASE2017 \nth{3} place \cite{3}  & 0.09$/$95.3 & 0.10$/$95.3 & 0.36$/$79.5 & 0.18$/$90.0 & 0.23$/$88.4 & 0.11$/$94.3 & 0.32$/$82.1 & 0.22$/$88.2\\
		\hline
		
		R-FCN  \cite{4}                   & 0.06$/$97.2 & 0.10$/$94.6 & 0.32$/$81.4 & 0.18$/$90.5 & $-\ /\ -$   & $-\ /\ -$   & $-\ /\ -$   & 0.32$/$82.0\\
		R-CRNN \cite{5}                   & 0.09$/\ -$    & 0.04$/\ -$    & 0.14$/\ -$    & 0.09$/$95.5 & $-\ /\ -$   & $-\ /\ -$   & $-\ /\ -$   & 0.23$/$87.9\\
		Multi-resolution \cite{8}         & 0.11$/$94.3 & 0.04$/$97.8 & 0.18$/$90.6 & 0.11$/$94.2 & 0.26$/$86.5 & 0.16$/$92.1 & 0.18$/$91.1 & 0.20$/$89.9\\
		Multi-scale \cite{54}             & 0.22$/$89.0 & 0.14$/$92.8 & 0.18$/$91.0 & 0.18$/$90.9 & $-\ /\ -$   & $-\ /\ -$   & $-/-$       & 0.33$/$83.9 \\
		T-F Attention \cite{43}           & 0.10$/$95.1 & 0.01$/$99.4 & 0.16$/$91.5 & 0.09$/$95.3 & 0.18$/$91.3 & \textbf{0.04}$/$\textbf{98.2} & 0.17$/$90.8 & 0.13$/$93.4\\
		\hline
		
		Our Baseline                      & 0.09$/$94.5 & 0.06$/$96.1 & 0.20$/$89.3 & 0.12$/$93.3 & 0.17$/$90.1 & 0.08$/$93.9 & 0.19$/$88.7 & 0.15$/$90.9\\
		AdaMD-LR                        & 0.07$/$96.0 & 0.05$/$95.7 & 0.18$/$90.2 & 0.10$/$94.0 & 0.14$/$91.4 & 0.07$/$95.2 & 0.13$/$92.1 & 0.11$/$92.9\\
		AdaMD-Default                   & 0.05$/$97.1 & 0.03$/$97.3 & 0.14$/$91.1 & 0.07$/$95.2 & 0.13$/$92.4 & 0.08$/$94.6 & 0.11$/$94.0 & 0.10$/$93.7\\
		AdaMD-Balanced                  & \textbf{0.04}$/$\textbf{98.2} & 0.02$/$98.8 & \textbf{0.12}$/$\textbf{92.5} & \textbf{0.06}$/$\textbf{97.6}
										   & \textbf{0.10}$/$\textbf{94.0} & 0.05$/$96.1 & \textbf{0.10}$/$\textbf{94.2} & \textbf{0.08}$/$\textbf{94.7}\\
		\hline

		\end{tabular}
\end{center}
\end{table*}

\begin{table*}[th]
\begin{center}
\caption{Results on DCASE 2016 Task 3 and DCASE 2017 Task 3}
\label{other_results}
		\begin{tabular}{c|c|c|c|c|c|c|c|c|c|c}
		\hline
		\multirow{2}*{Method}
			~ & \multicolumn{5}{c|}{\textbf{DCASE 2016 Task 3}} & \multicolumn{5}{c}{\textbf{DCASE 2017 Task 3}}\\
			\cline{2-11}
			~       & Baseline & \nth{1} \cite{81} & \nth{2} \cite{83} & \nth{3} \cite{84}  & AdaMD           & Baseline & \nth{1} \cite{82} & \nth{2} \cite{85}  & \nth{3} \cite{86}  & AdaMD\\
		\hline
		Error Rate  & 0.8773   & 0.8051        & 0.9056        & 0.9124         & \textbf{0.7821} & 0.9358 & 0.7914 & 0.8080 & 0.8251 & \textbf{0.7723} \\
		\hline
		F1-score    & 34.3\%   & 47.8\%        & 39.6\%        & 41.9\%         & \textbf{48.7\%} & 42.8\% & 41.7\% & 40.8\% & 39.6\% & \textbf{43.6\%} \\
		\hline
		\end{tabular}
\end{center}
\end{table*}

\section{Experiment Setting}

In this section, we introduce the datasets we use in our experiments. Then, we describe our data preprocessing method, evaluation metrics and baseline.

\subsection{Datasets}

The experiments in this paper were conducted on four datasets. The first one is DCASE 2017 Task 2 dataset, which is used to compare the performance of our algorithm with other state-of-the-art algorithms on monophonic AED task. In addition, we used DCASE 2016 Task 3 dataset and DCASE 2017 Task 3 dataset to compare our method with the results of the first three teams on polyphonic AED tasks. The last one is a self-made factory mechanical dataset, which is used to verify the application performance of our algorithm in the complex and noisy environment.

\subsubsection{DCASE 2017 Task 2} There are three types of events in this dataset: \textit{baby cry}, \textit{glass break} and \textit{gun shot}. Each type has about 60 source segments and 1000 background segments. In most cases, it is quite difficult to collect rare audio events in real life, therefore, the development set and evaluation set are created by artificial synthesizing.

For the development dataset, we generate 5000 samples for each category in training part and 500 samples for each category in testing part. The proportion of target events is set to be 0.99 in training part and 0.5 in testing part (in line with competition reports). There are 500 samples in each category in the official evaluation set of the competition. In order to maintain fairness, we use the same evaluation dataset in our experiments. The event-to-background ratio (EBR) is set to three values: -6dB, 0dB and 6bB. All audio files has 30-seconds length with 44,100 Hz and 24 bits.

\subsubsection{DCASE 2016 Task 3 and DCASE 2017 Task 3}

We use these two non-rare acoustic datasets to further evaluate our method. Both of these two datasets are polyphonic, which means several acoustic events will occur at the same time and the target is to predict the onset of all of them. As for the number of classes, DCASE 2016 Task 3 contains 18 classes and DCASE 2017 Task 3 contains 6 classes. These datasets will be harder than the previous one because a classification task is added besides the detection task.

\subsubsection{Factory Mechanical Dataset} We collected normal sounds of air compressor machines, as well as few audios when they have failures. The number of failure events is 134 and the number of normal recordings is 2000. We only use 34 event segments in the training dataset and the others are used in the testing dataset. We use the same tool as DCASE dataset to generate training and testing dataset. All audio files has 30-seconds length with 44,100 Hz and 24 bits. In order to analyze the model's ability against noise, we also conduct some experiments with added Gaussian noise to all generated audio segments during the training and test stage.

\subsection{Data Preprocessing and Feature Extraction}

We use \textit{librosa}\footnote{https://librosa.github.io/librosa/} to extract Mel-filter bank (f-bank) as the input of our neural network.
The frame length is 0.04s with a 0.02s stride.
The number of FFTs is set to 2048, and the number of Mel filters is set to 128.
Since the input dimension of neural networks is fixed, we need to cut the f-bank feature of one audio file into several samples before inputting the f-bank into the model, and the step size of the cutting process is $0.5\tau$. Then, we normalize the feature of each sample with the mean and variance normalization:
\begin{equation}
x_{out} = \frac{x_{in}-\mu(x_{in})}{\sigma(x_{in})}
\end{equation}

\subsection{Evaluation Metrics}

The evaluation metrics \footnote{\url{https://github.com/TUT-ARG/sed_eval}} we use are consistent with DCASE 2017 Task 3 \cite{59}: the correct criterion for event prediction is that onset position error does not exceed 500ms, and the offset is not calculated. Two event-based metrics are used in the experiments: error rate (ER) and F-1 score. Their formulations are based on true positives (TP), false positives (FP), and false negatives (FN):
\begin{equation}
P = \frac{TP}{TP+FP},\ \ R=\frac{TP}{TP+FN}
\end{equation}
\begin{equation}
ER = \frac{FN+FP}{N}, \ \ {F1}_{score} = \frac{2 \times PR}{P+R}
\end{equation}
where $N$ represents the total number of samples, $P$ and $R$ are called precision and recall respectively.

\subsection{Baseline}

The baseline provided by DCASE 2017 Task 2 only consists of fully connected layers with dropout, which cannot provide competitive results.
Therefore, we build a more powerful baseline based on CRNN structure that is very similar to the first place on DCASE 2017 Task 2.
The only difference is that we use the 2-dimensional convolution layer with 3*3 kernel rather than the 1-dimensional one.

\section{Experiment Discussion}

In this section, we first describe the experiments we conducted on DCASE 2017 Task 2, DCASE 2016 Task 3, and DCASE 2017 Task 3 datasets. After that, we analyze the function of each module in our framework with ablation experiments on DCASE 2017 Task 2. Finally, we carry out a verification experiment on the mechanical failure dataset to further validate our algorithm.

\subsection{Monophonic Detection}

Tab.~\ref{compare_table} shows the comparison with state-of-the-art methods on DCASE 2017 Task 2 dataset. Our final model is named as AdaMD-Balanced. First of all, we compare our AdaMD method with the top three teams in DCASE 2017 Challenge (first block). Our result exceeds the first place in development dataset except for \textit{glass break} event, and exceeds this team for all events in evaluation dataset. It is worth noting that all the teams participating in the competition adopt system ensemble method, that is, training multiple models for one event, then use the output of artificially weighted average of these models as the final prediction. Instead of using complex and time-consuming ensemble method, we report our results with a single model for each event.

In the second block, four other state-of-the-art algorithms are also compared, including FCN\cite{4}\cite{5}, multi-resolution\cite{8}, and attention mechanism\cite{43}. Among them, \cite{43} achieves quite promising results on \textit{glass break} event, but in terms of average performance, our algorithm is better than all of them in both ER and F1-score metrics.

Finally, we compare the AdaMD-Balanced model with three ablative models: the baseline model, the model using logistic regression as ensemble method (AdaMD-LR), and the model without threshold balance (AdaMD-Default). Here, the threshold balance refers to setting different thresholds for different events based on the results of development dataset (see Fig.~\ref{binary_threshold}). Opposite to threshold balance setting, the default setting means the thresholds of all three events are set to 0.5.
The comparison shows that our adaptive multi-scale method outperforms the model that uses simple ensemble method such as logistic regression. In addition, we also prove that setting different thresholds for different events will also improve the accuracy.

\subsection{Polyphonic Detection}

Tab.~\ref{other_results} shows the comparison results on DCASE 2016 Task 3 dataset and DCASE 2017 Task 3 dataset. Since these two datasets are designed for polyphonic detection, we modify the last fully-connected layer of our model to output the prediction for each class. Specifically, the dimension of the output for each frame equals to the number of event class and a sigmoid activation function is used to make sure the probability of the activity of each class falls in $[0, 1]$. All hyper-parameters keep the same and are shown in Tab.~\ref{parameter_table}. We compared our method with the official baseline and the first three teams in the challenge. According to the results, our AdaMD outperforms the first place in both datasets \cite{81}\cite{82}, proving the advantages of multi-scale model.

\begin{figure}[!t]
\centering
\includegraphics[width=8.5cm]{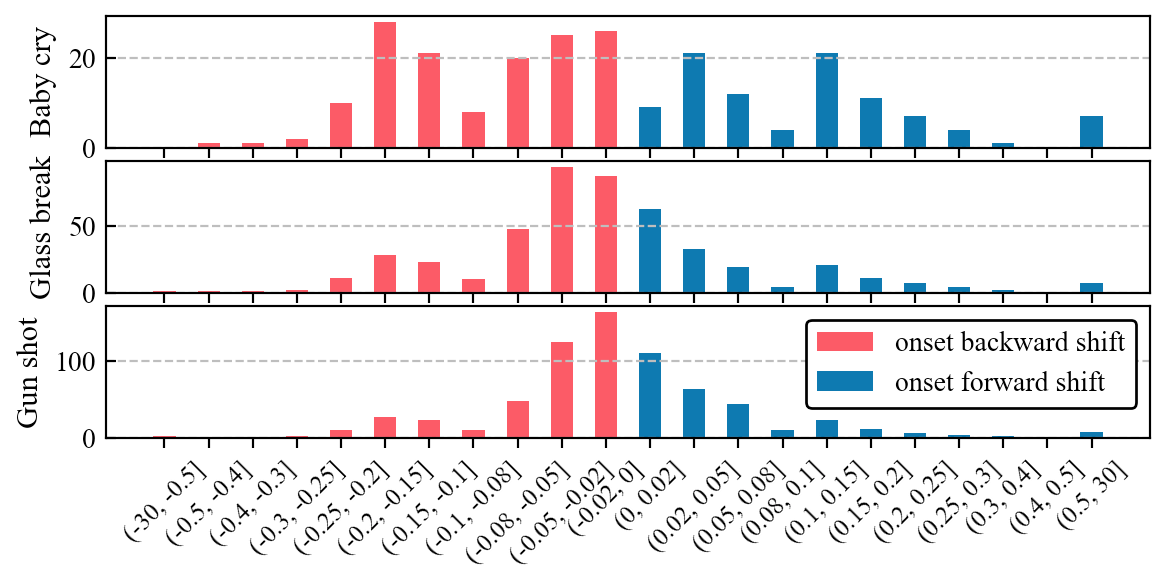}
\caption{Onset shift error of three events. The unit of x-axis is second. The shift time is divided into 24 ranges and all testing files fall in these ranges. The red bar and the blue bar respectively represent the samples that predict the onset later or earlier than groundtruth.}
\label{hist}
\end{figure}

\begin{figure}[!t]
\centering
\includegraphics[width=8.5cm]{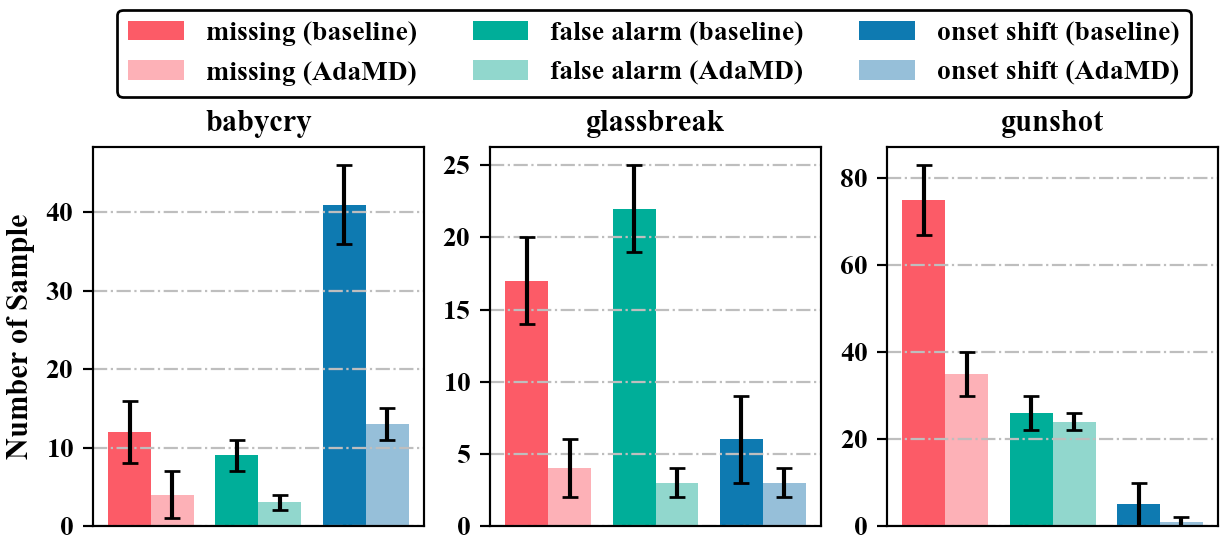}
\caption{Analysis of three types of error (missing, false alarm and onset shift). The transparent bar represents the result of AdaMD, while the solid bar represents the result of our baseline.}
\label{hist_three_types}
\end{figure}

\subsection{Discrepancy of Events}

We record the error reasons of all three events on DCASE 2017 Task 2, and show the analysis of the baseline and our AdaMD in Fig.~\ref{hist_three_types}. From the results of baseline, we can tell that the error type of the three events are different, which is in line with our analysis of acoustic events in Section \uppercase\expandafter{\romannumeral4}. The main error reason for \textit{baby cry} is that the prediction of onset is not accurate. This is because the high intra-class variation of the features, so this event can only be detected on a large scale. It is difficult to determine the specific location of events on a small scale. In contrast, onset prediction of \textit{glass break} is more accurate, and the rate of false alarm and missing is relatively low, because the characteristics of this event is more consistent, thus it belongs to a relatively easy task. Onset prediction of \textit{gun shot} is also accurate, but the main problem of \textit{gun shot} is the high rate of missing. Our analysis shows that the short duration of \textit{gun shot} and the inconsistency of temporal scale make this event easy to be ignored, thus the missing rate is much higher than others. However, as long as it is recognized, the accurate onset can be obtained.

The results of our adaptive multi-scale method show that all three events gain an improvement of onset accuracy and a decrease of missing rate. This confirms that our method can adaptively tackle different kinds of events.

\begin{figure}[!t]
\centering
\includegraphics[width=8cm]{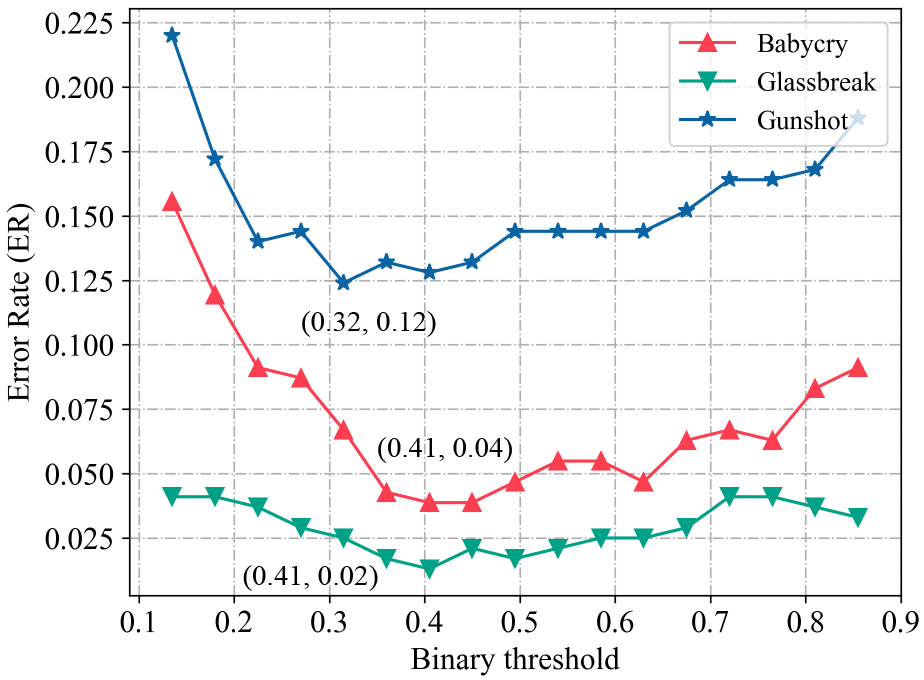}
\caption{Results of the binary threshold exploration. Error rates of three events are shown in different color and the notation in the figure represents the optimal threshold of corresponding event.}
\label{binary_threshold}
\end{figure}

\subsection{Impact of Threshold and Structure}

In our method, the output probability of neural network needs to be compared with a threshold to determine whether this frame is active or not, thus the selection of this threshold is very significant. In general, it is intuitive to set 0.5 as the default threshold, but the optimal values for different events could be different. Therefore, we conduct several experiments to explore the impact of this threshold on development dataset and show the results in Fig.~\ref{binary_threshold}.

Although the output of our AdaMD is continuous, the label we use for training is binary. That explains why we use a BCE as our loss function. Because no threshold is required during the training stage, we train one model and use different thresholds in testing stage to get the results. From Fig.~\ref{binary_threshold}, we know that for our AdaMD model, the optimal thresholds for these three events are 0.32, 0.41 and 0.41, respectively. We call the model using the optimal threshold as balanced model and the model using 0.5 threshold as default model.

\begin{figure*}[!t]
\centering
\includegraphics[width=18cm]{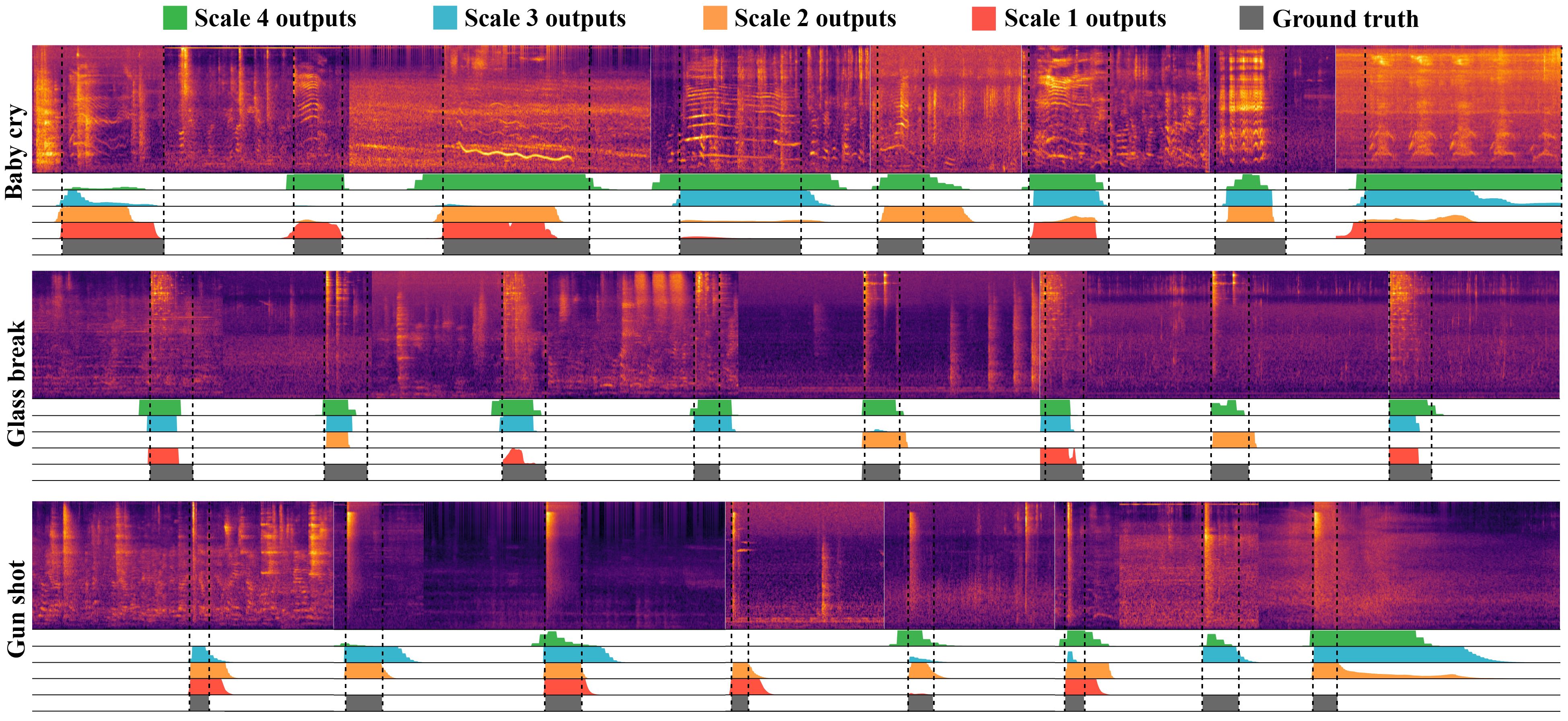}
\caption{Samples that are close to missing. Outputs of 4 different scales and the groundtruth are displayed in different colors. Since the event is quite short, we only select the positive segments and concatenate them into one long segment.}
\label{multiscale_missing}
\end{figure*}

\begin{figure*}[!t]
\centering
\includegraphics[width=18cm]{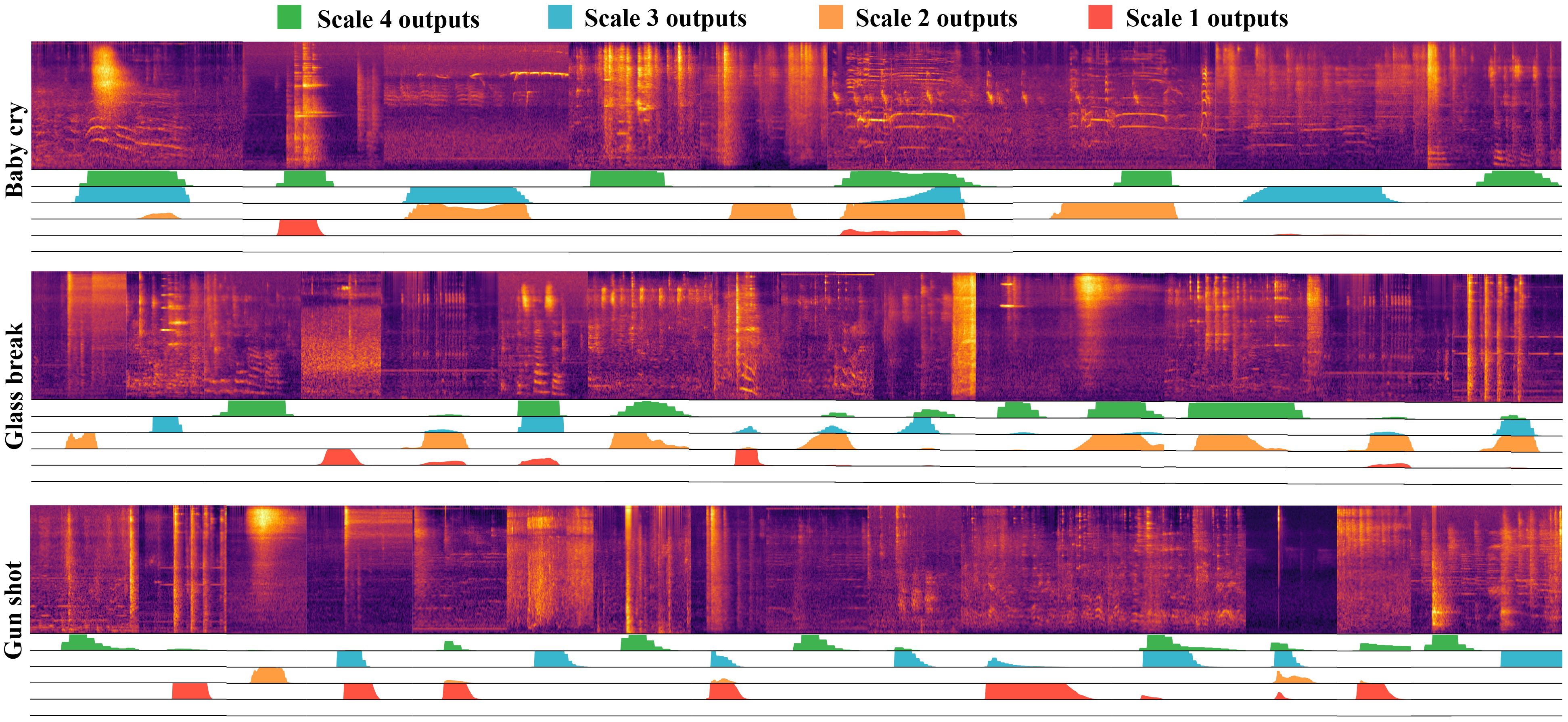}
\caption{Samples that are close to false alarm. Outputs of 4 different scales are displayed in different colors, and no groundtruth label is shown because there is no event in these segments. Since the event is quite short, we only select the positive segments and concatenate them into one long segment.}
\label{multiscale_false_alarm}
\end{figure*}

Another important part is the GRU module. the function of GRU and the specific settings of it are explored and shown in Tab.~\ref{rnn_table}. We draw several conclusions as follows:
\begin{enumerate}[]
	\item The existence of GRU module is quite necessary, since removing GRU module will lead to a poor result. This is mainly because the data is time sequence, which has correlations between frames, thus GRU can extract more information in the time domain.
	\item Bi-directional GRU is better than its uni-directional counterpart.
	Uni-directional GRU only considers the sequence information from front to back, which is of course in line with the law of nature. However, we can also extract features in the order from back to front, as the reverse processing of Filter-bank is still meaningful. The bi-directional approach provides us more temporal information.
	\item Multi-layer GRU is better than the single-layer one. The reason is that the feature extraction ability of single fully-connected layer is limited, while multi-layer structure has more capability of fitting non-linear functions.
\end{enumerate}

\subsection{Ablation for Multi-scale}

To explore why our model is better than the baseline, we first analyze the contribution of multi-scale network structure. In Fig.~\ref{multiscale_missing}, we show several typical examples of three events. In these examples, there are missing errors in one or two scales. By using our adaptive combination method, we can get the correct final prediction in most cases. However, if only one scale is considered, there will be a large number of missing errors. Similarly, in Fig.~\ref{multiscale_false_alarm}, we show some examples that one or two output scales have false alarm errors. However, enhanced by our method, the final output for most cases do not show false alarm after weighted output.

In Fig.~\ref{multiscale_missing} and Fig.~\ref{multiscale_false_alarm}, we also noticed that for one event, false alarm and missing errors often occur in several specific scales, while other scales output accurate results (e.g. gunshot has nearly no false alarm in scale 2). For different events, the most accurate scale is usually different, which supports that the characteristics of different events are usually expressed in different scales. In the final combined output, the more accurate scale will usually be assigned with the higher weight.

\begin{figure}[!t]
\centering
\includegraphics[width=8.5cm]{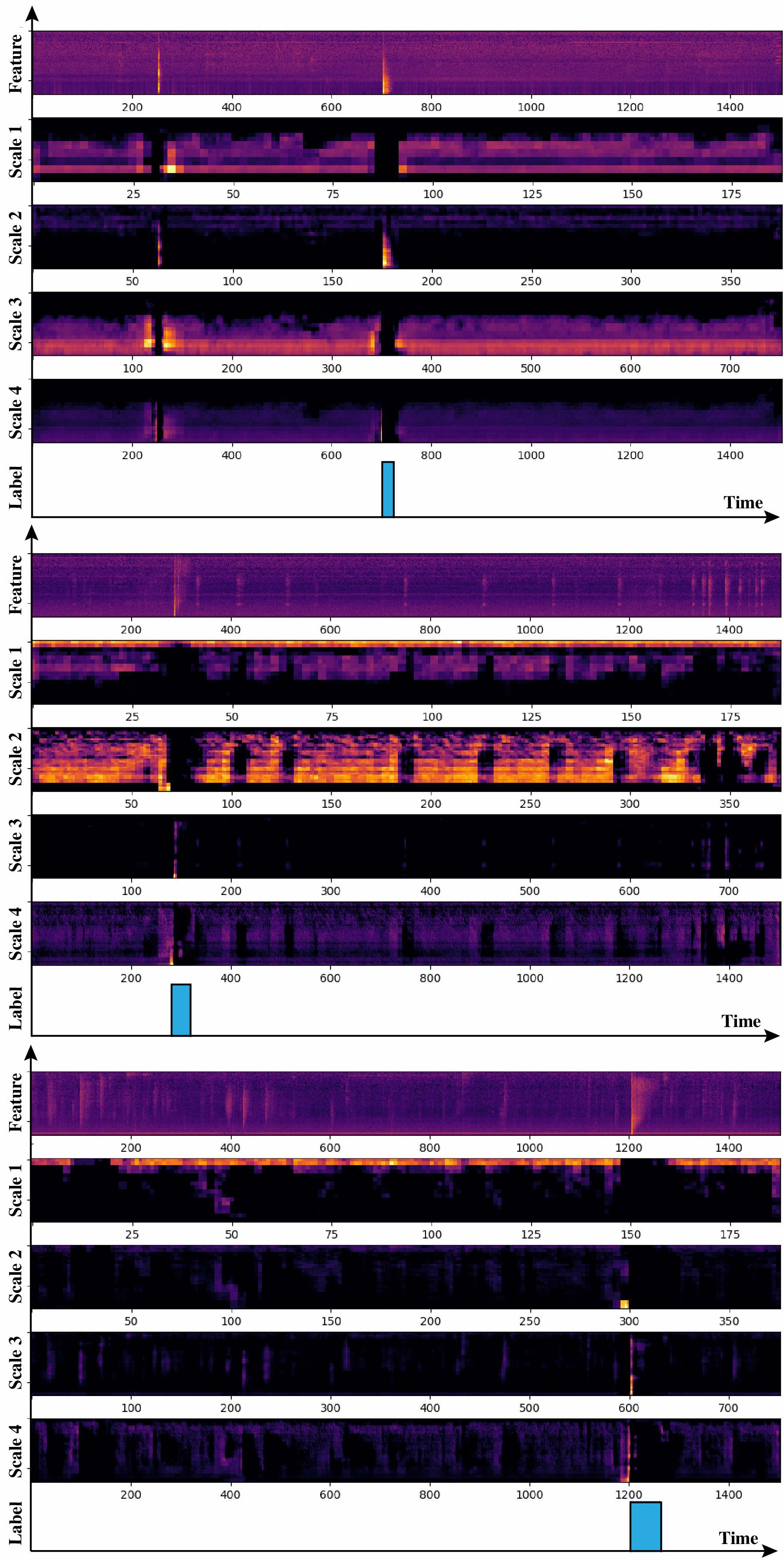}
\caption{Three examples of heatmap are displayed. We show the original feature, four heatmaps from different scales, and groundtruth label together. The label is frame-wise binary.}
\label{heatmap}
\end{figure}

In Fig.~\ref{heatmap}, we use three examples to show the contribution of different heatmaps, including the original feature, four heatmaps from different scales, and groundtruth label. The figure shows that different scales focus on different region, which supports the statement that multi-scale structure helps capture more information.

In addition to above qualitative analysis, we also provide some quantitative experiments. In Tab.~\ref{multi_table}, we compare the results of individual training of each scale with the results of average fusion for all scales. It can be concluded that no individual scale achieves the best output in all three events. While after average fusion, although the results of some events become worse, the overall prediction are better than those with only one scale.

\begin{table}[th]
\begin{center}
\caption{Influence of GRU Module (Error Rate)}
\label{rnn_table}
		\begin{tabular}{c|c|c|c|c|c}
		\hline
		Method & Layer & baby cry & glass break & gun shot & average\\
		\hline
		CNN only     & $-$ & 0.35 & 0.15 & 0.42 & 0.31\\
		w$/$ uni-GRU & 3   & 0.10 & 0.07 & 0.23 & 0.13\\
		w$/$ bi-GRU  & 1   & 0.06 & 0.05 & 0.14 & 0.08\\
		w$/$ bi-GRU  & 3   & \textbf{0.04} & \textbf{0.02} & \textbf{0.12} & \textbf{0.06}\\
		\hline
		\end{tabular}
\end{center}
\end{table}

\subsection{Ablation for Adaptive Training}

In this section, we conduct experiments to analyze the contribution of adaptive training to the entire algorithm. The results are shown in Tab.~\ref{adaptive_table}. According to the conclusion of the previous section, we find that each event has an optimal scale to get the best result, so we design a model that assigns a large weight $s_n^k$ to the optimal scale $k$ during training. For example, for \textit{gun shot} event, we set the weight of scale 1 to 10, and set the weights for other three scales to 1. This kind of model is named \textit{gun shot-weighted}. One thing to be noted is that the number 10 here is a hyper-parameter to ensure that the optimal branch contributes more than others. In Tab.~\ref{adaptive_table}, we also show single output of different scales obtained by adaptive training, which is used to compare with the single output of different scales without adaptive algorithm.

By comparing the experimental results, we find that enlarging the weight for optimal scale indeed improves the performance of an event. However, due to the discrepancy of acoustic events, artificially designing the weight for optimal branch would be time-consuming. In contrast, the adaptive training algorithm proposed by us automatically learns the best branch and assign a reasonable weight for it.

\begin{table}[th]
\begin{center}
\caption{Influence of Multi-scale Structure (Error Rate)}
\label{multi_table}
		\begin{tabular}{c|c|c|c|c}
		\hline
		Method  & baby cry & glass break & gun shot & average\\
		\hline
		scale 1 (no adaptive) & 0.11 & 0.07 & 0.18 & 0.12\\
		scale 2 (no adaptive) & 0.10 & 0.04 & 0.20 & 0.11\\
		scale 3 (no adaptive) & 0.09 & 0.05 & 0.17 & 0.10\\
		scale 4 (no adaptive) & 0.07 & 0.06 & 0.19 & 0.11\\
		\hline
		average (no adaptive) & \textbf{0.07} & \textbf{0.04} & \textbf{0.16} & \textbf{0.09}\\
		\hline
		\end{tabular}
\end{center}
\end{table}

\begin{table}[th]
\begin{center}
\caption{Influence of Adaptive Fusion (Error Rate)}
\label{adaptive_table}
		\begin{tabular}{c|c|c|c|c}
		\hline
		Method  & baby cry & glass break & gun shot & average\\
		\hline
		baby cry-weighted      & 0.06 & 0.08 & 0.18 & 0.10\\		glass break-weighted   & 0.11 & 0.04 & 0.20 & 0.11\\
		gun shot-weighted      & 0.12 & 0.05 & 0.14 & 0.10\\
		\hline
		scale 1 (adaptive) & 0.08 & 0.04 & 0.15 & 0.09\\
		scale 2 (adaptive) & 0.07 & 0.03 & 0.14 & 0.08\\
		scale 3 (adaptive) & 0.08 & 0.05 & 0.13 & 0.08\\
		scale 4 (adaptive) & 0.05 & 0.03 & 0.16 & 0.08\\
		\hline
		adaptive fusion & \textbf{0.04} & \textbf{0.02} & \textbf{0.12} & \textbf{0.06}\\
		\hline
		\end{tabular}
\end{center}
\end{table}

\subsection{Verification Experiment}

In this part, we conduct an experiment on our factory mechanical dataset to test the model's noise-resistant ability and sensitivity to the number of training sample. Since the factory environments are usually noisy, most acoustic features of machine are submerged in irrelevant sounds. Therefore, it is quite important for the model to have better noise-resistant capability. In our experiment setting, we artificially generate a Gaussian additive noise and add it to the acoustic signal after the noise being multiplied with an amplification factor. We change the value of this factor and record its influence to our model. Four different models are examined here, two of which are trained on clear dataset and tested on noisy dataset (noted as mismatched), while the other two are trained and tested on dataset with the same noise amplification factor (noted as matched).

\begin{figure}[!t]
\centering
\includegraphics[width=8.5cm]{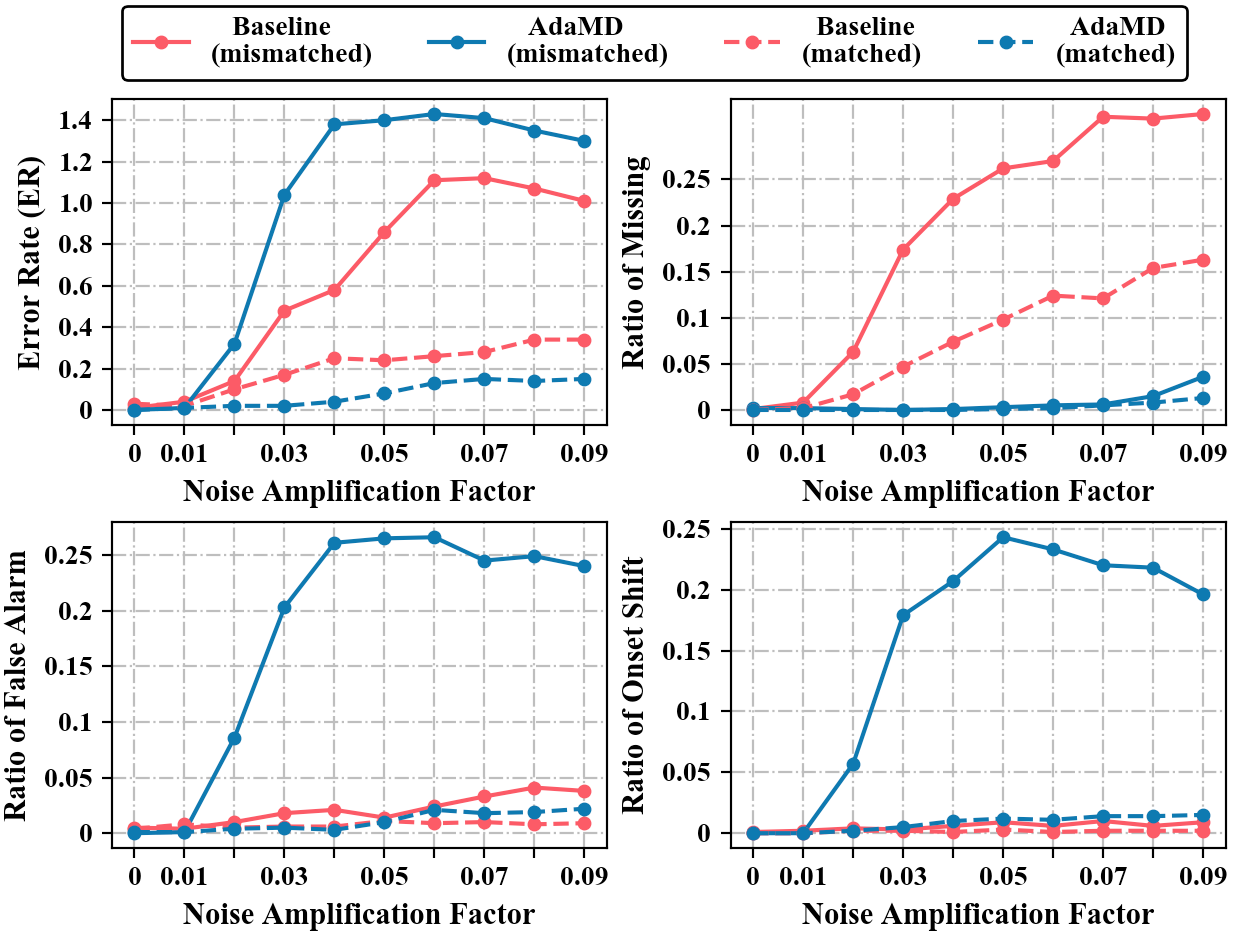}
\caption{Results of noise-resistant experiment. The top-left figure shows the results of error rate, and the rest three figures show the three different kinds of error respectively (missing, false alarm and onset shift).}
\label{noise}
\end{figure}

In Fig.~\ref{noise}, we show the results of both our AdaMD and baseline. For the matched models, the AdaMD obviously outperforms the baseline and it shows promising resistance to the additive noise. However, for the mismatched models, the AdaMD has larger error rate than the baseline when faced with large noise. One reasonable explanation is that the AdaMD is overfitting to the clean dataset, which means the mismatching between training and testing dataset will reduce the performance.

\begin{figure}[!t]
\centering
\includegraphics[width=8.5cm]{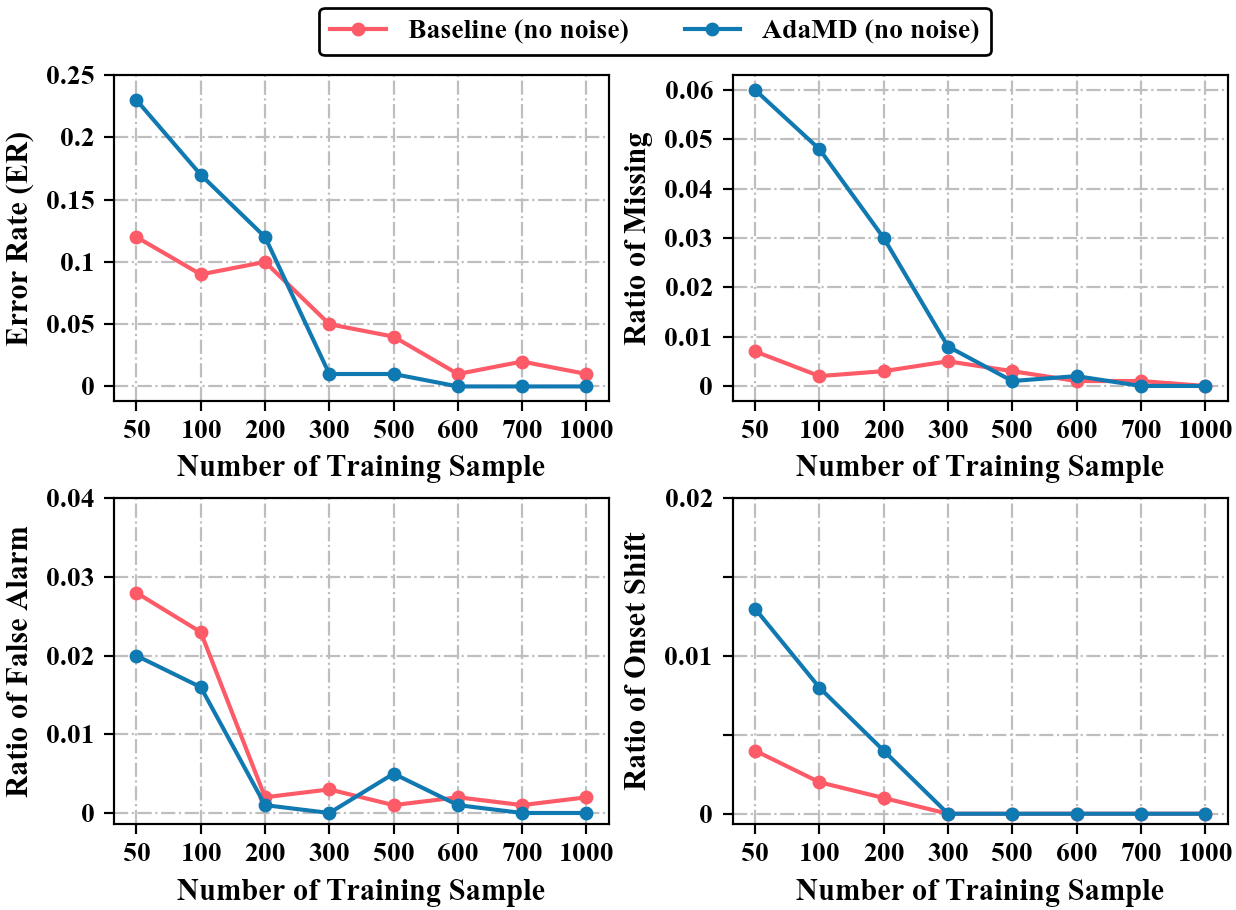}
\caption{Results of overfitting-analysis experiment. The top-left figure shows the results of error rate, and the rest three figures show the three different kinds of error respectively (missing, false alarm and onset shift).}
\label{sample}
\end{figure}

To explore more about the overfitting problem, we use another experiment with different numbers of training sample. In Fig.~\ref{sample}, we show the error rate of both AdaMD and baseline. When the training dataset is small, AdaMD indeed has more serious overfitting problem than the baseline, but their performances are nearly equal when a larger dataset is provided. We think this phenomenon is mainly because AdaMD has much more model parameters, which can be avoided when more data are available.

\section{Conclusion}

Detecting and locating acoustic events is an important event analysis method. Firstly, according to the characteristics of time and frequency domain, this paper provides an analysis for some acoustic events. Through this analysis, we claim that multi-scale feature extraction is important for the event detection, which guide us to develop an adaptive multi-scale event detection method. In this method, we build several weak classifiers at different temporal scales to extract feature and output prediction, separately. In order to achieve the best results for classifiers at each scale, we adaptively assign the weights of samples in the training process. Finally, the correct rate of each weak classifier is calculated based on the validation dataset, and the final fusion is carried out using this correct rate as the weight.

The experimental results on three standard challenge dataset show that our method has lower ER and higher F1-score on both monophonic and polyphonic situations than other algorithms. Through a number of ablation experiments, we analyze the contribution and impact of each module in the framework, proving the effectiveness of multi-scale and adaptive design. In order to verify the performance of our algorithm in the actual environment, we also utilize it to detect a variety of mechanical equipment failure events in the complex environment of the factory. The results affirm the capability of our algorithm to be used in real world.


\bibliographystyle{IEEEtran}
\bibliography{mybib}

\begin{IEEEbiography}{Wenhao Ding}
recevied the B.S. degree in electronic engineering from Tsinghua University, Beijing, China in 2018. He is now a Ph.D. student of Mechanical Engineering in Carnegie Mellon University, Pittsburgh, U.S. His research interests are rare event simulation and time sequence analysis, and he also focus on several application fields including acoustic event detection and speaker recognition.

\end{IEEEbiography}

\begin{IEEEbiography}{Liang He}
Liang HE received the B.S. degree in communication engineering from Civil Aviation University of China, Tianjin, China, in 2004, the M.S. degree in information science and electronic engineering from Zhejiang University, Hangzhou, China, in 2006, and the Ph.D. degree in information and communication engineering from Tsinghua University, Beijing, China, in 2011.
He is an Associate Professor at the Department of Electronic Engineering, Tsinghua University. His research interests are in the area of speech signal processing and artificial intelligence, primarily in speaker recognition, language recognition and acoustic event detection.
\end{IEEEbiography}

\end{document}